Title: Proper acceleration, geometric tachyon and dynamics of a
fundamental string near D$p$ branes
Authors: Ashok Das, Sudhakar Panda and Shibaji Roy
Comments: 30 pages, latex, to appear in Class. and Quant. Grav.  

We present a detailed analysis of our recent observation that the origin
of the geometric tachyon, which arises when a D$p$-brane propagates in the
vicinity of a stack of coincident NS5-branes, is due to the proper
acceleration generated by the background dilaton field. We show that when
a fundamental string (F-string), described by the Nambu-Goto action, is
moving in the background of a stack of coincident D$p$-branes, the geometric
tachyon mode can also appear since the overall conformal mode of the induced
metric for the string can act as a source for proper acceleration. We also
studied the detailed dynamics of the F-string as well as the instability by
mapping the Nambu-Goto action of the F-string to the tachyon effective action
of the non-BPS D-string. We qualitatively argue that the condensation of the
geometric tachyon is responsible for the (F,D$p$) bound state formation.

\documentclass[12pt]{article}
\usepackage{amsfonts}
\usepackage[centertags]{amsmath}
\usepackage{amssymb}
\usepackage{cite}
\usepackage{psfrag}
\usepackage{graphicx}
\usepackage{bbm,bm}
\usepackage{epsfig, multicol}

\allowdisplaybreaks
\begin{document}
\topmargin 0pt
\oddsidemargin 0mm
\def\be{\begin{equation}}
\def\ee{\end{equation}}
\def\bea{\begin{eqnarray}}
\def\eea{\end{eqnarray}}
\def\ba{\begin{array}}
\def\ea{\end{array}}
\def\ben{\begin{enumerate}}
\def\een{\end{enumerate}}
\def\nab{\bigtriangledown}
\def\tpi{\tilde\Phi}
\def\nnu{\nonumber}
\def\lll{\label}
 
\newcommand{\vs}[1]{\vspace{#1 mm}}
\newcommand{\dsl}{\pa \kern-0.5em /} 
\def\a{\alpha}
\def\b{\beta}
\def\g{\gamma}\def\G{\Gamma}
\def\d{\delta}\def\D{\Delta}
\def\ep{\epsilon}
\def\et{\eta}
\def\z{\zeta}
\def\t{\theta}\def\T{\Theta}
\def\l{\lambda}\def\L{\Lambda}
\def\m{\mu}
\def\f{\phi}\def\F{\Phi}
\def\n{\nu}
\def\p{\psi}\def\P{\Psi}
\def\r{\rho}
\def\s{\sigma}\def\S{\Sigma}
\def\ta{\tau}
\def\x{\chi}
\def\o{\omega}\def\O{\Omega}
\def\k{\kappa}
\def\pa {\partial}
\def\ov{\over}
\def\br{\\}
\def\ud{\underline}
%
\begin{center}
{\large{\bf Proper acceleration, geometric tachyon and dynamics of a 
fundamental string near D$p$ branes}}

\vs{6}

{Ashok Das$^{a,c}$\footnote{E-mail: das@pas.rochester.edu},
Sudhakar Panda$^b$\footnote{E-mail: panda@mri.ernet.in},
and Shibaji Roy$^c$\footnote{E-mail: shibaji.roy@saha.ac.in}}

\vspace{4mm}

{\em
$^a$ Department of Physics and Astronomy,\\
University of Rochester, Rochester, NY 14627-0171, USA\\

\vs{4}

$^b$ Harish-Chandra Research Institute\\
Chhatnag Road, Jhusi, Allahabad 211019, India\\

\vs{4}

$^c$ Saha Institute of Nuclear Physics\\
1/AF Bidhannagar, Calcutta 700064, India\\}

\end{center}

\begin{abstract}

We present a detailed analysis of our recent observation that the origin 
of the geometric tachyon, which arises when a D$p$-brane propagates in the 
vicinity of a stack of coincident NS5-branes, is due to the proper 
acceleration generated by the background dilaton field. We show that when 
a fundamental string (F-string), described by the Nambu-Goto action, is 
moving in the background of a stack of coincident D$p$-branes, the geometric 
tachyon mode can also appear since the overall conformal mode of the induced 
metric for the string can act as a source for proper acceleration. We also 
studied the detailed dynamics of the F-string as well as the instability by 
mapping the Nambu-Goto action of the F-string to the tachyon effective action 
of the non-BPS D-string. We qualitatively argue that the condensation of the 
geometric tachyon is responsible for the (F,D$p$) bound state formation. 

\end{abstract}

\newpage

\section{Introduction} 

Type II string theories are  known to admit two types of D$p$-branes 
as classical solutions. The BPS D$p$-branes 
\cite{Horowitz:1991cd,Duff:1994an,Polchinski:1995mt}
are supersymmetric and stable 
while the non-BPS D$p$ branes \cite{Sen:1999mg,Schwarz:1999vu}
are neither supersymmetric nor  stable. The 
instability of the non-BPS D$p$ branes arises because the lowest lying 
state of the open string theory with both ends attached to the brane 
corresponds to a tachyon. The dynamics of both the BPS as well as the 
non-BPS D$p$ branes are described by Dirac-Born-Infeld (DBI) actions 
involving the appropriate background fields. The DBI action for the non-BPS 
D$p$ brane \cite{Sen:1999md,Garousi:2000tr,Bergshoeff:2000dq,Kluson:2000iy} 
is a functional of the tachyon field and has a prefactor in the 
Lagrangian density that corresponds to the potential for the tachyon field. 
The tachyon potential has the interesting property that it has a finite 
maximum at the origin (in the field space) while it vanishes asymptotically 
leading to a vanishing pressure for late times \cite{Sen:2004nf}. 
The DBI action for the 
BPS D$p$ brane \cite{Leigh:1989jq,Schmidhuber:1996fy,Polchinski:1996na}, 
on the other hand, does not have any functional dependence 
on a tachyon degree of freedom and the multiplicative factor in the 
Lagrangian density depends on the dilaton background of the theory.

In spite of these differences, it has been noticed recently that in 
the vicinity of a stack of NS5-branes, the dynamics of a BPS D$p$ brane 
resembles that of a non-BPS D$p$ brane \cite{Kutasov:2004dj}. 
In fact, it has been shown that 
the DBI action for the radial mode living on the BPS D$p$-brane can be 
mapped to the tachyon field on the non-BPS D$p$-brane. In this case the 
exponential dilaton prefactor in the Lagrangian density of the BPS D$p$-brane 
can be thought of as playing the role of the effective potential for the 
tachyon. This potential falls off exponentially to zero at short distances 
while it behaves like a gravitational potential for large separations so 
that as the BPS D$p$-brane approaches the NS5-branes, this potential 
resembles the tachyon potential in the DBI action for the non-BPS D$p$-brane. 
Furthermore, the pressure for this system has also been observed to fall 
off exponentially to zero at late times, much like that in the case of the 
rolling tachyon in the decay of a non-BPS D$p$-brane to tachyon matter 
whose equation of state corresponds to that of a pressureless fluid
\cite{Sen:2002nu,Sen:2002in,Sen:2002an}. As a 
result, the radial mode of the BPS D$p$-brane near a stack of NS5-branes 
has been called a ``geometric tachyon". Such an instability can be physically 
understood because the configuration of the BPS D$p$-brane in the background 
of the stack of NS5-branes breaks all of the space-time supersymmetry of 
the system. 

This instability has been further studied by compactifying one of the 
transverse directions of the NS5-branes on a circle and placing the BPS 
D$p$-brane at a point diametrically opposite to the NS5-branes 
\cite{Kutasov:2004ct}. Such a 
configuration corresponds  to a saddle point in the potential energy of 
the BPS D$p$-brane leading to a tachyonic mode for translations along the 
compactified direction (circle). More recently a close relationship between 
the geometric tachyon and the universal open string tachyon has been 
extensively discussed by Sen \cite{Sen:2007cz} (see also
\cite{Israel:2007zc}, \cite{Kluson:2007hb}). The geometric tachyon 
has also played an 
interesting role in the inflationary models in cosmology 
\cite{Ghodsi:2004wn,Papantonopoulos:2006eg,Thomas:2005fu,
Panda:2005sg,Panigrahi:2007sq,Panigrahi:2008kg}.

In  a recent letter \cite{Das:2008af}, we studied the origin of the 
tachyonic instability 
in the motion of a BPS D$p$-brane in the vicinity of the stack of 
coincident NS5-branes in terms of the motion
of a point particle in the transverse space of the five-branes. We observed
that in this case the particle experiences a proper acceleration due to the
background dilaton field and as a result deviates from its geodesic. We 
argued that it is this dilaton dependent acceleration which is responsible 
for the motion of the radial mode
to be that of an inverted simple harmonic oscillator at small separations, 
leading to the tachyonic
instability. In this paper we give a detailed and systematic analysis 
of our results. 

We extend our observation to other dynamical systems involving different 
backgrounds. For example, instead of the BPS D$p$-brane, we consider the 
motion of a fundamental string (F-string) described by the Nambu-Goto 
action where there is no exponential dilaton prefactor. If the F-string 
is in the background of the NS5-branes,  we know that the system is 
supersymmetric (it is
S-dual to the D1-D5 system) and stable so that the F-string is nondynamical. 
On the other hand  when the fundamental string is in the background of 
a stack of coincident D$p$-branes, the 
dynamics of the F-string would still be described by the Nambu-Goto 
action without the dilaton prefactor. Naively, therefore, in this background  
one would not expect a proper acceleration associated with the motion of 
the F-string. However,  we know that the F-D5 system 
is S-dual to the D1-NS5
system and for the latter, as we have discussed the dynamics of the 
radial mode becomes
tachyonic near the NS5-branes. For this to be compatible with our 
proposal, therefore,  it is necessary that there should be a source of 
proper acceleration in the system. That this is indeed the case can be 
seen from the fact that, even though the Nambu-Goto action is free from 
the dilaton prefactor, it inherits one from the induced metric when the 
overall conformal factor in the background metric of the D5-brane is 
taken out. This conformal factor gives rise to the proper acceleration 
which is responsible for the deviation from geodesic motion leading to 
the tachyonic instability. As before, this conformal factor can be thought 
of as the tachyon potential in the open string
effective action for the tachyon. The same conclusion also follows 
for other D$p$-brane backgrounds with $p<5$.  

We  study in detail the dynamics of the F-string in various D$p$-brane 
backgrounds with $p\leq 5$. We map the Nambu-Goto action to the effective 
action for the tachyon in a non-BPS D-string. 
We show that when the F-string is far away from the
D$p$-brane,  the potential $V(T)$ (where $T$ denotes the tachyon field to be 
defined later in terms of the radial mode $R$) reduces to the expected 
attractive gravitational potential, but as it approaches the D$p$-brane 
the potential vanishes
exponentially for $p=5$ but as a power of $T$ for $p<5$. This fall off 
is the expected behavior of the effective potential for the tachyon
\cite{Sen:2004nf}. 
We solve the dynamical equation for the F-string and determine its 
trajectory in various cases of interest. We obtain the expression for 
pressure when the separation between the F-string and the D$p$ brane 
approaches zero and we show  that at late times, the pressure vanishes 
exponentially for $p=5$ and as a power when $p<5$, resulting in a 
pressureless dust (or tachyon matter) \cite{Sen:2002in}. This is 
indeed expected of an unstable
D$p$-brane with the tachyon rolling down to the minimum of the potential. 
We thus recover the  
analog of a rolling tachyon on an unstable D-string as the dynamics
of the F-string propagating in the D$p$-brane background. Of course, 
it would be interesting to know what would correspond to the end-point 
state of this tachyon condensation. From the tension formula
following from \cite{Lu:1999uca,Schwarz:1995dk}, we conjecture that as the 
F-string approaches the 
D$p$-branes, the F-string 
loses almost all of its energy density and thus the F-string really 
gets melted
down into the D$p$-branes forming an (F, D$p$) bound state. This would 
correspond to a
non-threshold bound state and would preserve half of the 
space-time supersymmetry.

This paper is organized as follows. In the next section, we briefly discuss
the motion of a point particle in a gravitational field subjected in addition to a proper
acceleration. We point out how a Lagrangian formulation can be used to obtain
the geodesic motion, modified by the acceleration term, as Euler-Lagrange
equation of motion. In section 3, we investigate the dynamics of a D$p$-brane
in the background of a stack of NS$5$-branes as a point particle motion in the
transverse space. This analysis is then extended to the motion of a
fundamental string in the background of a stack of coincident D$p$-branes in
section 4. We discuss the case of D5-branes in subsection 4.1 and other
D$p$ ($p<5$) branes in subsection 4.2. Here we  also
briefly discuss the limitations of the supergravity description as well as the
end-point of the geometric tachyon condensation. We present our
conclusions in section 5.

\section{Proper acceleration}

In this section, we review briefly the concept of proper acceleration in both
flat space time as well as in a curved background. To begin with let us
consider a relativistic point particle moving along a straight trajectory 
(linear motion) in flat space time with the line element
\begin{equation}
ds^{2} = - d\tau^{2} = \eta_{\mu\nu} dx^{\mu}dx^{\nu},\label{flatlineelement}
\end{equation}
where we are using the metric with signature $(-,+,\cdots,+)$. If the particle 
is moving along the $x$-axis, the Lorentz factor follows from 
\eqref{flatlineelement} to be (we are setting $c=1$)
\begin{equation}
\gamma = \frac{dt}{d\tau} = \frac{1}{\sqrt{1 - v^{2}}},\label{lorentzfactor}
\end{equation}
where
\begin{equation}
v = \frac{dx}{dt},\label{instantaneousvelocity}
\end{equation}
represents the instantaneous velocity of the particle along the $x$-axis. 
It follows from the definition in \eqref{lorentzfactor} that
\begin{equation}
\gamma^{2}v^{2} + 1 = \gamma^{2}.\label{identity}
\end{equation}
The motion of the particle in this case is effectively two dimensional 
and neglecting the other spatial coordinates we note that the proper 
velocity of the particle can be represented as the two dimensional vector
\begin{equation}
u^{\mu} = \frac{dx^{\mu}}{d\tau} = (\gamma, \gamma v),\label{propervelocity}
\end{equation}
which can be seen using \eqref{lorentzfactor} to satisfy
\begin{equation}
\eta_{\mu\nu} u^{\mu}u^{\nu} = -1.\label{speednormalization}
\end{equation}

If $v$ is not constant (namely, if there is an acceleration along the
$x$-axis), the motion of the particle is nonuniform and we can define 
the proper acceleration of the particle as
\begin{equation}
a^{\mu} = \frac{du^{\mu}}{d\tau} = \gamma\left(\frac{d\gamma}{dt}, 
\frac{d (\gamma v)}{dt}\right).\label{properacceleration}
\end{equation}
Using the identity in \eqref{identity} we can easily show that
\begin{equation}
\frac{d\gamma}{dt} = v\,\frac{d(\gamma v)}{dt} = \alpha v,\label{gammaevoln}
\end{equation}
so that we can write
\begin{equation}
a^{\mu} =   \frac{du^{\mu}}{d\tau} = \alpha 
(\gamma v, \gamma) =  \alpha \epsilon^{\mu\nu} u_{\nu},
\label{accelerationcomponents}
\end{equation}
where we have identified
\begin{equation}
\alpha = \frac{d (\gamma v)}{dt},\label{instantaneousacceleration}
\end{equation}
and it is clear from \eqref{accelerationcomponents} that $\alpha$ 
is simply the instantaneous acceleration of the particle. 
Furthermore, $\epsilon^{\mu\nu}$ denotes the two dimensional 
Levi-Civita tensor with $\epsilon^{01} =1$ and  
it now follows from \eqref{propervelocity} and 
\eqref{accelerationcomponents} that 
\begin{eqnarray}
\eta_{\mu\nu} u^{\mu} a^{\nu} & = & 0,\nonumber\\
\eta_{\mu\nu} a^{\mu} a^{\nu} & = & \alpha^{2},\label{spacelikeacceleration}
\end{eqnarray}
so that the proper acceleration is space-like and is orthogonal to the 
proper velocity. As a result, \eqref{speednormalization} continues to 
hold even when the particle is subjected to a proper acceleration. 

If we assume that the instantaneous acceleration $\alpha$ is a constant, 
the dynamical equations \eqref{properacceleration} can be solved leading 
to the fact that a constant proper acceleration leads to a hyperbolic motion 
for the particle (as opposed to parabolic motion in the nonrelativistic
case). This can be seen easily as follows. From  
\eqref{accelerationcomponents}, we note that for constant $\alpha$, we have
\begin{equation}
\frac{d^{2}u^{\mu}}{d\tau^{2}}  = \alpha^{2} u^{\mu},\label{secondordereqn}
\end{equation}
which already reflects the hyperbolic nature of particle motion. Explicitly, 
for constant acceleration, we see from \eqref{accelerationcomponents} that 
\begin{equation}
u^{\mu} - \alpha \epsilon^{\mu\nu} x_{\nu} = {\rm constant},\label{gammaconstraint}
\end{equation}
and the solutions of \eqref{secondordereqn} are given by 
\begin{eqnarray}
t (\tau) & = & \frac{1}{\alpha}\,\sinh (\alpha \tau).\nonumber\\
x (\tau) & = & \frac{1}{\alpha}\left(\cosh (\alpha \tau) - 1 
+ \alpha\, x_{0}\right),\label{soln}
\end{eqnarray}
where $x_{0} = x (\tau=0)$ and we have chosen the initial conditions 
$t(\tau=0)=0,v(\tau=0)=0$. The particle follows a hyperbolic space-like 
trajectory
\begin{equation}
- t^{2} + \left(x - x_{0} + \frac{1}{\alpha}\right)^{2} = \frac{1}{\alpha^{2}},
\end{equation}
with the asymptotes of the hyperbola determined by the value of $\alpha$.

The notion of a proper acceleration can be carried over to a curved background 
with metric $G_{\mu\nu}$  in a straightforward manner. Here, of course,
coordinates are not four vectors, but the proper velocity is a four 
vector which satisfies the condition 
\begin{equation}
G_{\mu\nu} u^{\mu}u^{\nu} = -1,\label{curvednormalization}
\end{equation}
The definition of the proper acceleration \eqref{properacceleration} 
in this case can be generalized covariantly  as
\begin{equation}
a^{\mu} = D_{\tau}u^{\mu} = \frac{du^{\mu}}{d\tau} + 
\Gamma^{\mu}_{\nu\lambda} u^{\nu}u^{\lambda},\label{curvedacceleration}
\end{equation}
where $\Gamma^{\mu}_{\nu\lambda}$ denotes the Christoffel symbol constructed 
from the metric $G_{\mu\nu}$ of the curved background. Using the metric compatibility 
condition $D_{\tau}G_{\mu\nu} = 0$, it follows easily from 
\eqref{curvedacceleration} that 
\begin{equation}
G_{\mu\nu}u^{\mu}a^{\nu} = 0,
\end{equation}
and the normalization condition \eqref{curvednormalization} 
continues to hold even in the presence of an acceleration.

Let us next see how the geodesic motion of a particle is modified due to the
presence of an acceleraton. There are several ways of incorporating 
acceleration into an action formalism. From the point of view of our 
discussions in the following sections, let us consider the action given by
\be
S = -\int dt f(x) \sqrt{-G_{\mu\nu} \frac{dx^\mu}{dt}
  \frac{dx^\nu}{dt}} = \int dt L = -\int d\tau f(x) \sqrt{-G_{\mu\nu} 
\frac{dx^\mu}{d\tau} \frac{dx^\nu}{d\tau}},\label{modifiedaction}
\ee
where $f(x)$ is a given function of the coordinates and $G_{\mu\nu}$ is the 
metric of the curved space-time. The line element in this case is given by
\be
-d\tau^2 = G_{\mu\nu} dx^\mu dx^\nu,
\ee
so that the Lorentz factor takes the form,
\be
\gamma = \frac{dt}{d\tau} = \frac{1}{\sqrt{-G_{\mu\nu} \dot x^\mu \dot x^\nu}},
\ee
where an `overdot' represents a derivative with respect to $t$. 
It now follows from the action \eqref{modifiedaction} that  
\bea
\frac{\partial L}{\partial \dot x^\mu} &=& f(x) 
G_{\mu\nu} \frac{dx^\mu}{d\tau}\\
\frac{\partial L}{\partial x^\mu} &=& \frac{1}{2} f(x) \partial_\mu
G_{\nu\lambda} \frac{dt}{d\tau}\frac{dx^\nu}{dt}\frac{dx^\lambda}{dt} -
\frac{d\tau}{dt}\partial_\mu f(x).
\eea
Therefore, the Euler-lagrange equation for the system takes the form
\be
\frac{d\tau}{dt}\left(\frac{d}{d\tau}\left(f(x) G_{\mu\nu} \frac{dx^\nu}{d\tau}
    \right) - \frac{1}{2} f(x) \partial_\mu G_{\nu\lambda}\frac{dx^\nu}{d\tau}
\frac{dx^\lambda}{d\tau} + \partial_\mu f(x)\right) = 0.
\ee
The above equation can be simplified leading to the geodesic equation with an
acceleration term of the form
\be
\frac{d^2x^\mu}{d\tau^2} + \Gamma^\mu_{\nu\lambda} \frac{dx^\nu}{d\tau}
\frac{dx^\lambda}{d\tau} = a^\mu,\label{2acceleratedeqn}
\ee
where
\bea
a^\mu &=& -\left(G^{\mu\nu} + \frac{dx^\mu}{d\tau}
  \frac{dx^\nu}{d\tau}\right) \partial_\nu\left(\ln f(x)\right),
\label{2acceleration}\\
\Gamma^\mu_{\nu\lambda} &=& \frac{1}{2} G^{\mu\rho}\left(\partial_\lambda
  G_{\rho\nu} + \partial_\nu G_{\rho\lambda} - \partial_\rho
  G_{\nu\lambda}\right).\label{2connection}
\eea
We see from \eqref{2acceleration} that by construction we have $G_{\mu\nu} (dx^\mu/d\tau) a^\nu = G_{\mu\nu}
u^\mu a^\nu = 0$, where we have used $G_{\mu\nu} u^\mu u^\nu = -1$, which
shows that the proper velocity and the proper acceleration are orthogonal to
each other. We see from this simple example that a coordinate 
dependent overall multiplicative factor in front of the Lagrangian 
for a point particle motion leads to a proper acceleration which 
modifies the trajectory of the particle from that of a geodesic. In the next
sections, where we consider the motion of a BPS D$p$-brane in the background 
of a stack of coincident NS5-branes and the motion of a fundamental
string in the background of a stack of coincident D$p$-branes, we will show
how the motion can be regarded as the motion of a relativistic point particle 
in the
transverse space of NS5-branes in the first case and the D$p$-branes in
the second case respectively. We will make use of our discussion in this
section there and show that the motion in those cases also will be subjected
to a proper acceleration orthogonal to the proper velocity and thereby
leading to a deviation from the geodetic motion of the particle. This in turn will be shown
to give rise to the tachyonic instability when the branes come close to
each other.

\section{D$p$-brane motion in NS5-brane background}

We now present a detailed analysis of the dynamics of a D$p$-brane propagating in
the background of a stack of $N$ number of coincident NS$5$-branes\footnote{
See \cite{Elitzur:2000pq,Pelc:2000kb,Ribault:2003sg} for some related earlier 
works and \cite{Lu:1998vh,Mitra:2000wr,Alishahiha:2000qv} for the (NS5,D$p$)
supergravity configurations.} by
formulating this as the motion of a relativistic point particle in the
transverse space of the five-branes. We take the five-branes to be stretched
in the directions ($x^1, x^2,....,x^5$) and denote the world volume directions
to be $x^{\bar\mu}, {\bar\mu}= 0,1,....5$. The transverse directions are
labeled by $x^m, m= 6,7,8,9.$. The D$p$-brane lies parallel to the NS5-branes
along the directions $x^\mu, \mu= 0, 1, 2,...,p$ with $p\leq 5$. Thus the
D$p$-brane is taken to be point like in the directions $x^m$. Note that if we
place the D$p$-brane at a finitely large distance $r=(x^mx^m)^{1/2}$, from the
coincident five-branes, it experiences an attractive force due to gravitational
and dilatonic interactions between them. Since at weak string coupling
($g_s$), the D$p$-brane is lighter than
the stack of NS5-branes, their mass being proportional to $1/g_s$ and
$N/g_s^2$ respectively, it will move towards the NS5-branes. This motion
depends upon the the details of the background fields created by the
NS5-branes. So we give below the
background metric, the dilaton and the field strength for the NSNS gauge field 
of a stack of coincident NS5-branes as \cite{Duff:1990wv,Callan:1991dj},
\bea
ds^2 &=& - d\tau^{2} = \eta_{\bar \mu \bar \nu} 
dx^{\bar \mu} dx^{\bar \nu} + G_{mn} dx^m dx^n,\nnu\\
G_{mn} & = & H(r)\delta_{mn},\nnu\\
e^{2(\phi-\phi_0)} &=& H(r) = 1 + \frac{N \ell_s^2}{r^2},\nnu\\
H_{mnp} &=& - \epsilon^q_{mnp} \partial_q \phi,\label{backgrounds}
\eea 
where $\eta_{\bar \mu \bar \nu} = {\rm diag}(-1,+1,+1,\ldots,+1)$, 
$H$ is the harmonic 
function in the transverse space describing the $N$
coincident NS5-branes, $\ell_s$ denotes the string length and the string
coupling is $g_s = e^{\phi_0}$.

We denote the world volume-coordinates of the D$p$-brane by $\xi^\mu,
\mu=0,1,..., p$. However, we can use the reparametrization invariance on the
world-volume and identify $\xi^\mu = x^\mu$. Since our interest lies in the
dynamics of the D$p$-brane in the transverse space of the five-branes, denoted
by ($x^6, x^7, x^8, x^9$), they will give rise to
the scalar fields, $X^m(\xi^\mu), m= 6,7,8,9$ on the world-volume of
the D$p$-branes. The
dynamics of the D$p$-brane is described by the dynamics of these scalar fields
governed by the DBI action
\be
S_p = -\tau_p\int d^{p+1} \xi e^{-(\phi-\phi_0)} \sqrt{-{\rm
    det}\left(G_{\mu\nu}+ B_{\mu\nu}\right)}, \label{dbiaction}
\ee
where $\tau_p$ is the tension of the D$p$-brane. $G_{\mu\nu}$ and $B_{\mu\nu}$
are the induced metric and the $B$-field on the world-volume, 
given by ($X^{\bar{\mu}} = x^{\bar{\mu}}$)
\bea
G_{\mu\nu} &=& \frac{\partial X^A}{\partial\xi^\mu} \frac{\partial
  X^B}{\partial\xi^\nu} G_{AB}(X),\nnu\\
B_{\mu\nu} &=& \frac{\partial X^A}{\partial\xi^\mu} \frac{\partial
  X^B}{\partial\xi^\nu} B_{AB}(X).\label{inducedmetric}
\eea
Here the indices $A,B=0,1,\ldots,9$ and $G_{AB}$ and $B_{AB}$ are the metric
and the $B$-field in ten dimensions, given in \eqref{backgrounds}. 
We are interested in
the case when the fields representing the position of the D$p$-brane $X^m$,
$m=6,\ldots,9$, depend only on time, $X^m = X^m(t)$. In this case, the action
\eqref{dbiaction} simplifies considerably as we get contribution from the
metric only (the antisymmetric field does not contribute) and takes the form,
\be
S_p = -\tau_p V_p\int dt\, e^{-(\phi-\phi_0)} \sqrt{1- G_{mn} 
\dot X^m \dot X^n},\label{particleaction}
\ee
where $V_p$ is the volume of the $p$-dimensional space in which the D$p$-brane
is stretched out. The dilaton field and the metric $G_{mn}$ 
are related to the harmonic
function as noted in \eqref{backgrounds}.

We rewrite the above action in a suggestive form as,
\bea
S_p &=& - \tau_p V_p \int dt e^{-\bar{\phi}} \sqrt{-G_{\bar{m}\bar{n}} 
\dot X^{\bar{m}} \dot X^{\bar{n}}}\nnu\\
&=& -\tau_p V_p \int d\tau e^{-\bar{\phi}} \sqrt{-G_{\bar{m}\bar{n}} 
\frac{dX^{\bar{m}}}{d\tau}
\frac{dX^{\bar{n}}}{d\tau}},\label{covariantparticleaction}
\eea
where $\bar{\phi} = \phi-\phi_0$, $G_{\bar{m}\bar{n}} = (-1,\, G_{mn})$ with
$\bar{m}$ and $\bar{n}$ taking values $\bar{m} = (0, m) =0,6,7,8,9$; 
$X^0 = t$ and $\tau$ is the proper time.  The action 
\eqref{covariantparticleaction} can be
thought of as describing the dynamics of a relativistic point 
particle in gravitational as well
as dilatonic backgrounds. We remark that the action \eqref{dbiaction} reduces 
to that of a relativistic point particle because we assumed $X^{\bar m}$ to 
depend only on time. In principle it could depend on other world-volume 
coordinates of the D$p$-brane and then the action can not be reduced to the
point particle action.  Note that $\bar{\phi}$ does not depend on 
$X^0$ and, therefore,  on time explicitly. However, since it depends 
upon $X^m$, which is a
function of time, it has an implicit time dependence. We note the similarity
of the action \eqref{covariantparticleaction} with \eqref{modifiedaction} 
when $e^{-\bar \phi}$ is identified with $f$. The Lorentz 
factor here takes the form, 
\be
\gamma = \frac{dt}{d\tau} = \frac{1}{\sqrt{-G_{\bar{m}\bar{n}} 
\dot X^{\bar{m}} \dot X^{\bar{n}}}},\label{Dplorentzfactor}
\ee
Denoting the proper velocity of the particle as $u^{\bar{m}} = 
dX^{\bar{m}}/d\tau$, it can be checked easily that $G_{\bar{m}\bar{n}} 
u^{\bar{m}} u^{\bar{n}} = -1$ as in \eqref{curvednormalization}. As we 
mentioned in the introduction, in the vicinity of the NS5-branes, the dynamics
of a BPS D$p$-brane resembles that of a non-BPS D$p$-brane and there is a
tachyonic instability for the BPS D$p$-brane. However, we note that the 
momentum of the particle as obtained 
from \eqref{covariantparticleaction} is given as $P_{\bar{m}} = \tau_{p}V_{p} 
e^{-\bar{\phi}} G_{\bar{m}\bar{n}} u^{\bar{n}}$, which leads to $P^{2} = 
G^{\bar{m}\bar{n}}P_{\bar{m}}P_{\bar{n}} = - (\tau_{p}V_{p})^{2}
e^{-2\bar{\phi}}$ 
and this makes it clear that such a particle is not tachyonic in the 
conventional sense unless the dilaton field becomes complex.

Therefore, to understand the origin of the tachyonic instability, 
let us analyze the equations of motion following from 
\eqref{covariantparticleaction} which correspond to the motion of a 
particle in a curved background subject to an acceleration, namely, 
\be
\frac{d^2 X^{\bar{m}}}{d\tau^2} + \Gamma_{\bar{n}\bar{p}}^{\bar{m}}\, 
\frac{d X^{\bar{n}}}{d\tau} \frac{d
  X^{\bar{p}}}{d\tau} = a^{\bar{m}},\label{eqns}
\ee
where $\Gamma^{\bar{m}}_{\bar{n}\bar{p}}$ is the Christoffel symbol
constructed 
from the metric $G_{\bar{m}\bar{n}}$ and
and the proper acceleration $a^{\bar{m}}$ can be computed from 
eq.\eqref{2acceleration}
with $f=e^{-\bar \phi}$ as,
\be
a^{\bar{m}} = \left(G^{\bar{m}\bar{n}} + \frac{dX^{\bar{m}}}
{d\tau}\frac{dX^{\bar{n}}}{d\tau}\right)
\partial_{\bar{n}}\bar{\phi}.\label{acceleration}
\ee
Thus we note that the dilaton background is responsible for a 
proper acceleration leading to a deviation of the trajectory of the
particle from its geodesic. It can be checked easily that  $G_{\bar{m}\bar{n}}
u^{\bar{m}} a^{\bar{n}} = 0$ 
so that the proper acceleration is orthogonal to the proper velocity as would
be expected for a relativistic system. 
This is reminiscent of a Rindler particle executing hyperbolic motion 
\cite{Rindler}  and
clarifies the origin of the 
hyperbolic solution obtained in \cite{Kutasov:2004dj}. 

We can compute  the energy-momentum tensor by using
the general formula
\be
T^{\mu\nu} = - \frac{\partial L}{\partial(\partial_\mu X^m)}
\partial^\nu X^m + \eta^{\mu\nu} L,\label{tmunu}
\ee
where $L$ is the Lagrangian  of the action \eqref{covariantparticleaction} 
and the non-vanishing 
components take the explicit forms,
\bea
T^{00} &=& \tau_p V_p \gamma e^{-\bar{\phi}} \equiv E,\nnu\\
T^{ij} &=& -\tau_p V_p \gamma^{-1} e^{-\bar{\phi}} \delta^{ij} \equiv p
\delta^{ij},\label{energypressure} 
\eea
where $\gamma$ is the Lorentz factor defined in \eqref{Dplorentzfactor} 
and $E, p$ denote the
energy and the pressure of the system respectively. From  time translation 
invariance we expect energy to be conserved and similarly rotational 
invariance in the transverse space leads to the conservation of angular 
momentum in the system. 

The time component ($\bar{m}=0$) of the equation of motion \eqref{eqns} yields
\be
\frac{d\gamma}{dt} = \gamma \frac{d\bar{\phi}}{dt},\label{energyconservation}
\ee
which we recognize from \eqref{energypressure} to lead to 
conservation of energy. On the other hand,  using 
 \eqref{energyconservation} the dynamical equation  
\eqref{eqns} for $\bar{m}=m$  takes the form
\be
\ddot X^m + \Gamma^m_{\bar{n}\bar{p}} \dot X^{\bar{n}} 
\dot X^{\bar{p}} + G_{\bar{p}\bar{q}} \dot X^{\bar{p}} \dot
  X^{\bar{q}}  G^{mn} \partial_n\bar{\phi} = 0.\label{meqn}
\ee
We note that for large separations the leading behaviour of this 
equation is the free
particle motion described by $\ddot X^m = 0$. This corresponds to the
vanishing of the gravitational force as well as the  
acceleration $a^m$ for large separations in the leading order. We study 
below the dynamics of the system in the next to leading order.   

For this purpose it is simpler to work in the
spherical-polar coordinates. Using the fact that angular momentum is 
conserved, we can restrict the motion of the particle to a
plane with the radial mode $R$ and the angular mode $\Theta$. In this case
the line element in \eqref{backgrounds} takes the form 
\be
- d\tau^2 = - dt^2 + H dR^2 + H R^2 d\Theta^2,\label{sphericallineelement}
\ee
and correspondingly the Lorentz factor \eqref{Dplorentzfactor} becomes,
\be
\gamma = \frac{1}{\sqrt{1-H\dot R^2 - H R^2 \dot \Theta^2}}.
\label{sphericallorentzfactor}
\ee
The nonvanishing components of $\Gamma$ are given by,
\bea
\Gamma^R_{RR} &=& \partial_R \ln \sqrt{H},\quad 
\Gamma_{\Theta
  \Theta}^R = -R^{2} \partial_R\ln \sqrt{H R^2},\nnu\\
\Gamma^{\Theta}_{R \Theta} &=& \Gamma^{\Theta}_{\Theta R} =  \partial_R \ln
\sqrt{H R^2 }.\label{sphericalconnection}
\eea
In the spherical coordinates \eqref{meqn} has the form
\bea
\ddot R - R \dot \Theta^2 + \frac{1}{2 H^2} \partial_R H \left(2 H \dot R^2 -
  1\right) &=& 0,\nnu\\
\ddot \Theta + \frac{1}{H R^2} \partial_R \left(H R^2 \right) \dot \Theta \dot
R &=& 0.\label{sphericaleqns}
\eea
Since $\Theta$ is an angular coordinate, its conjugate gives the angular
momentum of the form  
\be
L = \tau_p V_p \gamma e^{-\bar{\phi}} H R^2 \dot \Theta = E H R^2 
\dot \Theta.\label{angmom}
\ee
Defining the quantity
\be
H R^2 \dot \Theta = \frac{L}{E} = \ell,\label{ell}
\ee  
we note that the second equation in \eqref{sphericaleqns} leads to 
the conservation condition
\be
\frac{d}{dt} (H R^2 \dot \Theta) = \frac{d}{dt} \left(\frac{L}{E}\right) 
= \frac{d\ell}{dt}= 0,\label{angmomconservation}
\ee
namely,  the angular momentum associated with the motion of the particle 
is conserved. 

The true dynamics of the system is contained in the $R$-equation in
\eqref{sphericaleqns}. 
Using \eqref{angmom}  as well as the fact that energy in 
\eqref{energypressure} is conserved, 
we obtain from \eqref{sphericallorentzfactor}
\be
\dot{R}^{2} = \frac{1}{H}\Big(1 -
\frac{1}{H}\Big(\Big(\frac{\tau_{p}V_{p}}{E}\Big)^{2} + 
\frac{\ell^{2}}{R^{2}}\Big)\Big) \geq 0,\label{bounds}
\ee
which determines (using the form of $H$ in \eqref{backgrounds}) that for
$\frac{\tau_{p}V_{p}}{E}\geq 1$, 
we must have $(N\ell_{s}^{2}-\ell^{2})\geq 0$ and $R^{2} \leq R_{0}^{2}$,
while for 
$\frac{\tau_{p}V_{p}}{E}\leq 1$, we can have either
$(N\ell_{s}^{2}-\ell^{2})\geq 0$ without any restriction on $R$, 
or $(N\ell_{s}^{2} - \ell^{2})\leq 0$ with $R^{2} \geq R_{0}^{2}$ where
$R_{0}^{2} = 
\left|\frac{N\ell_{s}^{2}-\ell^{2}}{\big(\frac{\tau_{p}V_{p}}{E}\big)^{2}-1}
\right|$. Since 
we are interested in the behavior of the system close to the origin $R\simeq 
0$, 
it is clear that we must have $(N\ell_{s}^{2} - \ell^{2})\geq 0$ independent 
of the value of the ratio $\frac{\tau_{p}V_{p}}{E}$. 
As it 
is natural to assume that the D$p$-brane starts out infinitely far 
away, we would assume $\frac{\tau_{p}V_{p}}{E}\leq 1$, although what 
is really important for the analysis of the behavior near the NS5 
branes is that $N\ell_{s}^{2}-\ell^{2}\geq 0$.

Using the angular-momentum conservation relation 
\eqref{angmomconservation}, the radial equation in \eqref{sphericaleqns} 
can be simplified to have the form,
\be
\ddot R - \frac{(N\ell_s^2-\ell^2) N\ell_s^2}{(R^2+N\ell_s^2)^3}\left(R +
  \left(\frac{2}{R_0^2} - \frac{1}{N\ell_s^2}\right)R^3\right) = 0 
\label{ddotR}
\ee
This is an exact expression and we will see the behavior of the radial
mode at the two extremes when $R\gg \sqrt{N} \ell_s$ as well as when $R\ll
\sqrt{N} \ell_s$ by expanding $(R^2 + N \ell_s^2)^3$. In the first case,
when $R\gg \sqrt{N} \ell_s$, we have
\be 
\ddot R + (N \ell_s^2 - \ell^2)N \ell_s^2\left(\frac{1}{N \ell_s^2} -
  \frac{2}{R_0^2}\right)\frac{1}{R^3} + O\left(\frac{1}{R^5}\right) = 0 
\label{ddotRg}
\ee
On the other hand for $R \ll \sqrt{N} \ell_s$, we have,
\bea
\ddot R &-& \frac{(N \ell_s^2 - \ell^2)}{(N \ell_s^2)^2} R + \frac{2(N
    \ell_s^2 - \ell^2)}{(N \ell_s^2)^3} \left(2 -
    \frac{N \ell_s^2}{R_0^2}\right) R^3\nnu\\ 
&-& \frac{3(N \ell_s^2 - \ell^2)}{(N \ell_s^2)^4}
  \left(3 - \frac{2 N \ell_s^2}{R_0^2}\right) R^5 + O(R^7) = 0 
\label{ddotRl}
\eea
Thus we note from \eqref{ddotRg} that when the D$p$-brane is far away from the
stack of NS5-branes, the associated particle moves in an attractive $-1/R^2$
potential (since $(N \ell_s^2 - \ell^2) > 0$) in four spatial dimensions 
(transverse directions to the NS5-branes) as expected. On the other hand,
when the D$p$ brane is at a distance $\sqrt{N} \ell_s$ away from the stack of
NS5-branes, the behavior of the radial mode becomes much more transparent if
we make a coordinate transformation $Z=1/R$. In this coordinate we can rewrite
the eq.\eqref{ddotRl} as,
\be
\ddot Z - \frac{(N \ell_s^2 - \ell^2)}{(N \ell_s^2)^{\frac{5}{2}}}
\left(\sqrt{N} \ell_s Z -\left(3 - \frac{2 N \ell_s^2}{R_0^2}\right)
  \frac{1}{(\sqrt{N} \ell_s Z)^3} + O\left(\frac{1}{(\sqrt{N} \ell_s
        Z)^5}\right) \right) = 0 \label{ddotZ}
\ee
We thus find that for $R\ll \sqrt{N} \ell_s$, i.e., $\sqrt{N} \ell_s Z \gg 1$
the above equation reduces, to the leading order,
\be
\ddot Z - \frac{1}{(N \ell_s^2)^2}\left(N \ell_s^2 - \ell^2\right) Z = 0
\label{simpleharmonic}
\ee
For $(N \ell_s^2-\ell^2)>0$, which is the case of our interest, we recognize
the above equation to correspond to the inverted simple harmonic oscillator. 
In the absence of the acceleration ($a^R = 0$), it can be easily checked that 
the radial equation for small $R$ reduces to $\ddot Z = 0$.  
The origin of the tachyonic instability is now clear, namely, it is the
acceleration due to the dilatonic background which is the source 
of the instability. Even though $P^{2}< 0$ indicating that the particle 
is not tachyonic in 
the conventional sense, the dilatonic force that it experiences in the 
background of the NS5 branes leads to hyperbolic motion 
and  the tachyonic instability in the system. We can solve
eq.\eqref{simpleharmonic} and the solution is given as (for $Z\sqrt{N} \ell_s
\gg 1$)
\be
Z = \frac{1}{R} = \frac{\tau_p V_p}{E \sqrt{N \ell_s^2 - \ell^2}}
\cosh\left(\frac{\sqrt{N \ell_s^2 - \ell^2}}{N \ell_s^2} t\right) \label{soln1}
\ee
Using this solution, we can simplify the expression for the pressure
given in \eqref{energypressure} as 
\be
p=-\tau_p V_p \gamma^{-1} e^{-\bar \phi}
= -\frac{E (N \ell_s^2 - \ell^2)}{N \ell_s^2} 
{\rm exp}\left(-2\frac{\sqrt{N \ell_s^2
- \ell^2}}{N \ell_s^2} t\right)\label{pressure}
\ee
So, the pressure falls off exponentially to zero
at late times, indicating that the system evolves into the
tachyon matter state at late times as for the case of rolling
tachyon on non-BPS branes. We note here that although  
\eqref{simpleharmonic} exhibits a tachyonic instability,  in order to
see where the instability occurs we look at the complete radial equation
\eqref{ddotR}. From here we identify the potential $V(R)$ in which the
particle moves by equating $\ddot R = -dV(R)/dR$. 
It is now easy to see from \eqref{ddotR} that the potential has a maximum at
$R=0$ (note that for $R_0 < \sqrt{N} \ell_s$, the case in which we are
interested, this is the only extremum) and so this is the point of
instability. Now we can also calculate the effective mass squared  of the
particle and it has the form, 
\be
m^{2} \equiv \left.\frac{d^2 V(R)}{dR^2}\right|_{R=0} = - 
\frac{(N\ell_{s}^{2} - \ell^{2})}{(N\ell_{s}^{2})^{2}}.\label{tachyonmass}
\ee
As the mass squared is negative, the particle effectively behaves as a
tachyon in the vicinity of NS5-branes.
However, it should be noted that since at $R=0$, the effective string coupling
$e^{\bar \phi} = H^{1/2}$ blows up, a full quantum string theoretic treatment
is necessary to identify the true position of the instability.

\section{Dynamics of F-string in the background 
of D$p$-branes}

We next investigate the motion of a fundamental string in type II string 
theories in the background of a stack of N coincident D$p$-branes 
(for $p\leq 5$). However, since the behavior of the F-string is quite
different when the background is D5-branes from those of D$p$-branes, with 
$p<5$ we will treat these two cases separately. First we will consider the
D5-brane background and then discuss about the other D$p$-brane backgrounds.

\subsection{F-string in D5-brane background}

The analysis of this system will follow closely that in the previous section. 
We recall that a well separated system of a fundamental string and a stack of 
parallel D$p$-branes leads to a complete breakdown of the space-time 
supersymmetries. Furthermore, as the F-string mass 
($\sim$ constant) is much less than the D$p$-brane mass ($\sim N/g_s$), where
$N$ is the number of D$p$-branes, the 
F-string will move towards the D$p$-branes at small string coupling because
of their gravitational and dilatonic interactions. 

The background fields describing the supergravity solutions of the coincident 
D5-branes have the forms \cite{Duff:1994an}
\bea
ds^{2} & = & H^{-\frac{1}{2}} \left(-dt^{2} + \sum_{i=1}^{5} 
(dx^{i})^{2}\right) + H^{\frac{1}{2}}\left(dr^{2} + r^{2} d\Omega_{3}^{2}
\right),\nonumber\\
e^{2(\phi -\phi_{0})} & = & H^{-1},\nonumber\\
H (r) & = & 1 + \frac{Ng_{s}\ell_{s}^{2}}{r^{2}},\nonumber\\
F_{7} & = & dH^{-1}\wedge dt\wedge dx^{1}\wedge\cdots \wedge dx^{5}.
\label{d5background}
\eea
Here we have assumed that $(t, x^{1},\cdots ,x^{5})$ denote the coordinates 
along which the D5-branes are spread while $(x^{6},\cdots ,x^{9})$ represent 
the transverse coordinates and $r= \sqrt{x^{m}x^{m}}$. $H(r)$ defines the 
harmonic function associated with the D5-branes and $F_{7}$ represents the 
7-form field strength.

The Nambu-Goto action describing the dynamics of the bosonic sector of a 
fundamental string in the background of the D5-branes is given by
\be
S_{F} = - \frac{1}{2\pi \ell_{s}^{2}} \int d^{2}\xi\, 
\sqrt{- \det G_{\mu\nu}},\label{fstring}
\ee
with $\mu,\nu=0,1$ correspond to the string world sheet indices while 
$\xi^{0},\xi^{1}$ denote the coordinates on the string world sheet. The 
induced metric on the world sheet of the fundamental string due to the 
background D5-branes defined in \eqref{d5background} is given by
\be
G_{\mu\nu} = \frac{\partial X^{A}}{\partial \xi^{\mu}}\frac{\partial X^{B}}
{\partial \xi^{\nu}}\,G_{AB},\label{fstringinducedmetric}
\ee
with $A,B=0,1,\cdots ,9$. As before, $X$'s denote the scalar fields associated
with the coordinates $x$'s. We can use the reparameterization invariance of 
the string world sheet to choose $\xi^{0}=x^{0}=t$ and $\xi^{1}=x^{1}$ so 
that the string is extended along $x^{1}$.

Let us note that there is no dilaton factor in the Nambu-Goto action 
\eqref{fstring} as opposed to the DBI action for the D$p$-brane in 
\eqref{dbiaction}. However, using \eqref{d5background} and 
\eqref{fstringinducedmetric} we can rewrite
\be
\sqrt{-\det G_{\mu\nu}} = \frac{1}{\sqrt H} \sqrt{1 + \eta^{\mu\nu} 
G_{mn} \partial_{\mu}X^{m} \partial_{\nu}X^{n}},
\label{sqrt}
\ee
where $\eta_{\mu\nu} = {\rm diag} (-1,1)$, and we have defined 
$G_{mn} = H (r) \delta_{mn}$ and $m,n=6,7,8,9$ describe the transverse space 
of the D5-branes. Using \eqref{sqrt}, the Nambu-Goto action \eqref{fstring} 
can be written as
\be
S_{F} = - \frac{1}{2\pi\ell_{s}^{2}} \int dt dx^{1}\,\frac{1}{\sqrt H}
\sqrt{1 + \eta^{\mu\nu}G_{mn} \partial_{\mu}X^{m}\partial_{\nu}X^{n}}.
\label{fstring1}
\ee

As in the previous section, we will be interested in the case when the 
scalar fields on the F-string depend only on time, namely, $X^{m} = 
X^{m} (t)$. In this case, \eqref{fstring1} simplifies and can be written as
\bea
S_{F} & = & - \tau_{F}V_{F} \int dt\, \frac{1}{\sqrt H}\, 
\sqrt{- G_{\bar{m}\bar{n}} \dot{X}^{\bar{m}} \dot{X}^{\bar{n}}}\nonumber\\
& = & - \tau_{F} V_{F} \int d\tau\, \frac{1}{\sqrt H}\,
\sqrt{- G_{\bar{m}\bar{n}} \frac{dX^{\bar{m}}}{d\tau} 
\frac{dX^{\bar{n}}}{d\tau}},\label{fstringfinal}
\eea
where we have identified $\tau_{F} = \frac{1}{2\pi\ell_{s}^{2}}$, $V_{F} = 
\int dx^{1}$ and $G_{\bar{m}\bar{n}}$ is defined in the last section. 
Comparing \eqref{fstringfinal} with \eqref{covariantparticleaction} we 
observe that the overall multiplicative factor $1/\sqrt{H}$ in the 
Lagrangian in \eqref{fstringfinal} can be thought of as playing the role of 
$e^{-\bar{\phi}}$ in the present case. Namely,  this factor can be thought of 
effectively as producing the acceleration in the present case much like the 
exponential dilaton prefactor was the source of acceleration for particle 
motion in the last section (see \eqref{modifiedaction}).

The dynamics of the fundamental string in the background of a stack of 
$N$ coincident D5-branes now follows from \eqref{fstringfinal} (see 
also \eqref{2acceleratedeqn} and \eqref{2acceleration}) 
\be
\frac{d^2 X^{\bar{m}}}{d\tau^2} + \Gamma_{\bar{n}\bar{p}}^{\bar{m}}\, 
\frac{d X^{\bar{n}}}{d\tau} \frac{d X^{\bar{p}}}{d\tau} = 
a^{\bar{m}},\label{fstringeqns}
\ee
with the acceleration in the present case given by
\be
a^{\bar m} = \left(G^{\bar m \bar n} + \frac{d X^{\bar m}}{d \tau}
\frac{d X^{\bar n}}{d \tau}\right) \frac{\partial_{\bar n} H}{2H}
\label{d5acceleration}
\ee
The energy and the pressure follows from the general form of the 
energy-momemtum tensor given in \eqref{tmunu} with the Lagrangian 
following from the action \eqref{fstringfinal} and have the forms,
\bea
T^{00} &=& \tau_F V_F \gamma H^{-\frac{1}{2}} = E\nnu\\
T^{ij} &=& - \tau_F V_F \gamma^{-1} H^{-\frac{1}{2}} \delta^{ij} = p 
\delta^{ij} \label{energypress}
\eea
where the Lorentz factor `$\gamma$' has the form given in 
\eqref{Dplorentzfactor}. The time
component ($\bar m =0$) of the equation of motion \eqref{fstringeqns} 
gives the energy
conservation relation $d\ln(\gamma H^{-1/2})/dt = 0$ whereas $\bar m = m$
gives,
\be
\ddot X^{\bar m} + \Gamma^m_{\bar n \bar p} \dot X^{\bar n} \dot X^{\bar p}
+ G_{\bar p \bar q} \dot X^{\bar p} \dot X^{\bar q} G^{mn} \frac{\partial_n H}
{2H} = 0 \label{ddotXnew}
\ee
The above equation being a second order differential equation, the motion
of the particle can be restricted to a plane. As before, introducing the
spherical-polar coordinates, the radial and the angular modes can be shown,
from \eqref{ddotXnew}, to satisfy exactly the same equations as given in
\eqref{sphericaleqns}. The $\Theta$ equation yields the conservation of
angular momentum as given in \eqref{angmomconservation} and the radial 
equation has the form
\be
\ddot R - \frac{(Ng_s\ell_s^2-\ell^2) Ng_s\ell_s^2}{(R^2+Ng_s\ell_s^2)^3}
\left(R +
  \left(\frac{2}{R_0^2} - \frac{1}{Ng_s\ell_s^2}\right)R^3\right) = 0
\label{ddotRF}
\ee
This is exactly the same equation we obtained for the motion of the
relativistic point particle corresponding to the motion of D$p$-branes in
NS5-brane background with $\ell_s^2$ replaced by $g_s\ell_s^2$. The reason for
this is that the configuration of F-string in D5-brane background is S-dual
to the configuration of D-string in NS5-brane background and under S-duality
$\ell_s^2$ changes to $g_s\ell_s^2$. It now obvious that the dynamics in 
this case is analogous to that of D-string in NS5-brane background we studied 
in the previous section. The radial equations in the two limits 
$R \gg \sqrt{N g_s} \ell_s$ and $R \ll \sqrt{N g_s} \ell_s$ have the same
forms as in \eqref{ddotRg} and \eqref{ddotRl} with $\ell_s^2$ 
replaced by $g_s\ell_s^2$.
Thus the whole analysis of the previous section goes through in a completely
parallel manner leading to the fact that in the vicinity of the stack of
D5-branes, the motion of F-string, described by a relativistic point particle,
is given by that of an inverted simple harmonic oscillator leading to the
tachyonic instability.
 
We would like to point out here that the motion of a D3-brane, described by 
the  DBI action, in the background of the D5-branes,  is given by the 
present analysis as well. This is because, for the D3-brane, the dilaton 
field satisfies $e^{2\phi} =1$ so that there is no dilaton prefactor in the 
DBI action. However, the overall multiplicative factor $H^{-\frac{1}{2}}$
would again arise from the induced metric as discussed in this section and 
this would be the source of proper acceleration. Therefore, a geometric 
tachyon would also arise for this system near the D5-branes. The dynamics
of a D$p$-brane in the background of a D$p'$-brane has been studied in
\cite{Panigrahi:2004qr}.

The tachyonic behavior of F-string in the vicinity of a stack of D5-branes
can also be understood, following \cite{Kutasov:2004dj}, by mapping the 
Nambu-Goto action
\eqref{fstring1} to the tachyon effective action of the non-BPS branes. For
this purpose, let us assume that in \eqref{fstring1} only the radial mode
$R=\sqrt{X^m X^m}$ has been excited and the angular modes are kept fixed, then
the action reduces to,
\be
S_F = - \frac{1}{2\pi \ell_s^2} \int dt dx^1 \frac{1}{\sqrt {H(R)}}
\sqrt{1 + \eta^{\mu\nu} H(R) \partial_\mu R \partial_\nu R} \label{tach1}
\ee
Now comparing this with the tachyon effective action of the non-BPS D-string
\be
S_{\rm nonBPS} = - \int dt dx^1 V(T) \sqrt{1 + \eta^{\mu\nu} \partial_\mu T
\partial_\nu T} \label{tach2}
\ee
where $T$ is the open string tachyon living on the non-BPS D-string and $V(T)$
is the tachyon potential. Now \eqref{tach1} can be mapped to \eqref{tach2}
if we identify
\bea
\frac{dT}{dR} &=& \sqrt{H(R)} = \sqrt{1 + \frac{N g_s \ell_s^2}{R^2}} 
\label{TR}\\
{\rm and} \quad V(T) &=& \frac{1}{2\pi\ell_s^2} \frac{1}{\sqrt {H(R)}}
= \frac{\tau_F}{\sqrt {H(R)}}\label{tachpotential}
\eea
The solution of eq.\eqref{TR} has the form,
\be
T(R) = \frac{1}{2} \sqrt{N g_s} \ell_s \ln\frac{\sqrt{R^2 + N g_s \ell_s^2}
- \sqrt{N g_s} \ell_s}{\sqrt{R^2 + N g_s \ell_s^2} + \sqrt{N g_s} \ell_s}
+ \sqrt{N g_s \ell_s^2 + R^2} \label{TRsoln}
\ee
upto an unimportant additive constant which we set to zero. From 
\eqref{TRsoln}, we find that as $R \to 0$,
\be
T(R \to 0) \to \sqrt{N g_s} \ell_s \ln\frac{R}{\sqrt{N g_s} \ell_s} \to
- \infty \label{asymp1}
\ee
and as $R \to \infty$,
\be
T(R \to \infty) \to R \to \infty \label{asymp2}
\ee
Using these relations \eqref{asymp1} and \eqref{asymp2}, we can find out
the behavior of the tachyon potential \eqref{tachpotential}
at the two extremes which we give below.
\bea
{\rm As,}\quad R &\to& \infty, \qquad V(T(R)) \to \tau_F
\left(1 - \frac{N g_s \ell_s^2}{2R^2}
\right)\,\, \simeq\,\, \tau_F \left(1 - \frac{N g_s \ell_s^2}{2T^2}\right)
\label{Vasymp1}\\
{\rm as,}\quad R &\to& 0, \qquad V(T(R)) \to \tau_F R\,\, \simeq\,\, 
\tau_F \sqrt{N g_s} \ell_s 
{\rm exp}\left(\frac{T}{\sqrt{N g_s} \ell_s}\right)\label{Vasymp2}
\eea 
This shows that when the F-string is far away ($R \gg \sqrt{N g_s} l_s$)
from the stack of D5-branes, the potential goes as $-1/T^2 \simeq -1/R^2$
which is the attractive gravitational potential in the transverse four 
dimensional space as expected and is consistent with our previous observation
\eqref{ddotRl}. On the other hand when the F-string comes closer 
($R \ll \sqrt{N g_s} \ell_s$) to the stack of D5-branes, $T \to -\infty$ and
the potential vanishes exponentially. This is what is expected of a tachyon
potential on the non-BPS branes. This behavior of the potential leads to the
two crucial properties \cite{Sen:2002an} of a tachyon effective action, 
namely, the absence of
the plane wave solution around the minimum of the potential (this is also 
related to our observation that near the D5-branes, the motion of the F-string,
described by a relativistic point particle, is given by the inverted simple
harmonic oscillator) and the exponential fall off of the pressure at late 
times. In order to understand the exponential fall off of pressure, we need
to solve the radial equation of motion. We have already given the solution
for D-string in NS5-brane background in \eqref{soln1}. The solution for
the motion of F-string in D5-brane background can be obtained from there
by replacing $\tau_1 V_1$ by $\tau_F V_F$ and $\ell_s^2$ by $\ell_s^2 g_s$.
So, the solution in this case would be,
\be
Z = \frac{1}{R} = \frac{\tau_F V_F}{E \sqrt{N g_s\ell_s^2 - \ell^2}}
\cosh\left(\frac{\sqrt{N g_s\ell_s^2 - \ell^2}}{N g_s\ell_s^2} t\right) 
\label{soln2}
\ee 
Therefore, the pressure given in \eqref{energypress} can be shown to behave as,
\be
p = -\frac{E (Ng_s \ell_s^2 - \ell^2)}{Ng_s \ell_s^2}
{\rm exp}\left(-2\frac{\sqrt{Ng_s \ell_s^2
- \ell^2}}{Ng_s \ell_s^2} t\right)\label{press}
\ee
This shows that the pressure indeed falls off exponentially to zero at late
times. However, we would like to remark here that the trajectory we have given
in \eqref{soln2} for $R \ll \sqrt{Ng_s}\ell_s$ is not sensible at all times.
The reason is that the supergravity configuration of a stack of D5-branes
given in \eqref{d5background} has a curvature singularity at $R=0$ (or $t \to
\infty$). In fact in order for the supergravity description
to remain valid the curvature must be small compared to the string
scale and so we have,
\be
{\cal R } = (\sqrt{N g_s} \ell_s R)^{-\frac{1}{2}} \ll \frac{1}{\ell_s} 
\label{curvature}
\ee
This implies that $R \gg \ell_s/(\sqrt{N g_s})$. So, the range in which the
trajectory \eqref{soln2} remains valid is
\be
\frac{\ell_s}{\sqrt{N g_s}} \ll R \ll \sqrt{N g_s} \ell_s \label{limit}
\ee
This can be satisfied for large $N$ and then the solution \eqref{soln2}
will remain valid in the above region \eqref{limit} only.

\subsection{F-string in D$p$-brane ($p<5$) backgrounds}

Let us next consider the motion of an F-string in the background of a stack
of coincident D$p$-branes (for $1 \leq p < 5$). The generalities of the
dynamics will remain exactly the same as in the previous subsection and we
will follow closely to that discussion. The dynamics of the F-string will
again be given by the Nambu-Goto action \eqref{fstring}. However, since now 
the background is different, the induced metric $G_{\mu\nu}$ will be 
different. The background fields describing the supergravity solution of
coincident D$p$-branes have the forms \cite{Duff:1994an},
\bea
ds^2 &=& H^{-\frac{1}{2}} \left(-dt^2 + \sum_{i=1}^p (dx^i)^2\right) 
+ H^{\frac{1}{2}} \left(dr^2 + r^2 d\Omega_{8-p}^2\right) 
\nonumber\\
e^{2(\phi-\phi_0)} &=& H^{\frac{3-p}{2}}\nonumber\\
F_{[p+2]} &=& dH^{-1}\wedge dt\wedge dx^1\wedge \ldots \wedge dx^p
\label{sugrasoln}
\eea
where $\mu,\nu = 0,1,\ldots,p$ label the coordinates along the D$p$-branes
and $m,n=p+1,\ldots,9$ label the coordinates transverse to the
D$p$-branes. $H$ is a harmonic function in the transverse space given by 
\be
H = 1 + \frac{d_p N g_s l_s^{7-p}}{r^{7-p}} = 1 + \frac{Q_p}{r^{7-p}}
\label{harmonicfn}
\ee
where $Q_p= d_p N g_s l_s^{7-p}$ and $d_p$ is a $p$-dependent constant 
given as $d_p = 2^{5-p} \pi^{(5-p)/2}
\Gamma((7-p)/2)$, $N$ is the number of D$p$-branes and $F_{[p+2]}$ is the
field-strength of an RR $(p+1)$-form gauge field that couples to the 
D$p$-branes. When $p=3$, the corresponding field-strength would be self-dual.

The Nambu-Goto action \eqref{fstring} in this background would take the 
same form as \eqref{fstring1} with $G_{mn} = H(r) \delta_{mn}$ and $H(r)$
as given in \eqref{harmonicfn}. When the scalar fields on the F-string
world-sheet
depend only on time, namely, $X^m = X^m(t)$, the Nambu-Goto action would
reduce to \eqref{fstringfinal} giving rise to the same equation of motion
\eqref{fstringeqns} with the proper acceleration given in 
\eqref{d5acceleration}. The time component of the equation of motion would give
the energy conservation relation with the energy and the pressure having the
same form as in \eqref{energypress}, whereas, the spatial component of the
equation of motion would be the same as \eqref{ddotXnew}. This being a 
second order
differential equation we can restrict the motion in a plane as before, then the
radial and the angular mode would give rise to the equations 
\eqref{sphericaleqns} obtained earlier. The $\Theta$ equations would give
the conservation of angular momentum \eqref{angmomconservation} and the
radial equation gives,
\bea
\ddot R && - \frac{1}{(R^{7-p} + Q_p)^3}\left(\frac{7-p}{2} Q_p^2 R^{6-p}
- (6-p) Q_p \ell^2 R^{2(6-p)-1}\right.\nnu\\
 \qquad\qquad\qquad && \left. - (7-p)\left(\frac{1}{2} + \frac{Q_p}
{R_0^{7-p}} - \frac{\ell^2}{R_0^2}\right)Q_p R^{2(6-p)+1} + \ell^2 R^{3(6-p)}
\right) = 0\label{ddotRp}
\eea
where $R_0$ is given by the equation
\be
\left(\frac{\tau_F V_F}{E}\right)^2 = 1 - \frac{\ell^2}{R_0^2} + \frac{Q_p}
{R_0^{7-p}}
\ee
Note that \eqref{ddotRp} is an exact equation of the radial mode $R$. When
the F-string is far away ($R \gg Q_p^{\frac{1}{7-p}}$) from the stack of 
D$p$-branes, eq.\eqref{ddotRp} gives,
\be
\ddot R - \frac{\ell^2}{R^3} + (7-p) Q_p \left(\frac{1}{2} + \frac{Q_p}
{R_0^{7-p}} - \frac{\ell^2}{R_0^2}\right)\frac{1}{R^{8-p}} = 0
\label{ddotradg}
\ee
Thus we note that for the vanishing angular momentum ($\ell=0$)
the particle motion is governed by an attractive $-1/R^{7-p}$ gravitational
potential as expected. On the other hand, when F-string is close 
($R \ll Q_p^{\frac{1}{7-p}}$) to the stack of D$p$-branes we use the full
radial equation \eqref{ddotRp} to identify the potential $V(R)$ in which 
the particle moves by equating $\ddot R = -dV(R)/dR$. From \eqref{ddotRp}
it is clear that the potential has an extremum at $R=0$ and this is the only
extremum when F-string is very close to the stack of D$p$-branes. Computing the
second derivative at $R=0$, we find that $d^2V/dR^2|_{R=0} = 0$. So, this
is a point of inflexion, indicating that there is an instability at this 
point. However, we note that except for $p=3$, in all other cases, either
the effective string coupling $e^{\bar \phi}$ blows up (for $p=1,2$) or the
curvature blows up (for $p=4$) at $R=0$ and so, we need full string theory 
to find the true position of the instability.

Now as in the previous subsection we can try to understand the instability
of the motion of F-string in tne vicinity of the stack of D$p$-branes, by
mapping the Nambu-Goto action \eqref{tach1} to the non-BPS D-string action
\eqref{tach2}. In the present case the mapping can be done if we identify,
\bea
\frac{dT}{dR} &=& \sqrt{H(R)} = \sqrt{1 + \frac{Q_p}{R^{7-p}}}
\label{TR1}\\
{\rm and} \quad V(T) &=& \frac{1}{2\pi\ell_s^2} \frac{1}{\sqrt {H(R)}}
= \frac{\tau_F}{\sqrt {H(R)}}\label{tachpotential1}
\eea
The solution of \eqref{TR1} is given as
\be
T(R) = - \frac{2R}{5-p}\left(1+ \frac{Q_p}{R^{7-p}}\right)^{\frac{1}{2}}
+ \frac{2(7-p)}{(9-p)(5-p)}\frac{R^{\frac{9-p}{2}}}{Q_p^{\frac{1}{2}}}
F\left(\frac{9-p}{2(7-p)},\frac{1}{2};1+\frac{9-p}{2(7-p)};-\frac{R^{7-p}}
{Q_p}\right)\label{solTR}
\ee
where $F(a,b;c;-R^{7-p}/Q_p)$ is a hypergeometric function. Now from
\eqref{solTR} we find that as $R \to 0$, $F(a,b;c;-R^{7-p}/Q_p) \to 1$
and so,
\be
T(R \to 0) \to - \frac{2 Q_p^{\frac{1}{2}}}{5-p} R^{-\frac{5-p}{2}} \to
-\infty\label{asymptotic1}
\ee
On the other hand as $R \to \infty$, the hypergeometric function behaves as
\be
F\left(a,b;c;-\frac{R^{7-p}}{Q_p}\right) \to \frac{\Gamma(c)\Gamma(b-a)}
{\Gamma(b)\Gamma(c-a)}\left(\frac{Q_p}{R^{7-p}}\right)^a + 
\frac{\Gamma(c)\Gamma(a-b)}
{\Gamma(a)\Gamma(c-b)}\left(\frac{Q_p}{R^{7-p}}\right)^b\label{hypproperty}
\ee
provided $(a-b)$ is not an integer. Using \eqref{hypproperty} we find from
\eqref{solTR} that in this case
\be
T(R \to \infty) \to R \to \infty\label{asymptotic2}
\ee
Using the relations \eqref{asymptotic1} and \eqref{asymptotic2} we analyse
the behavior of the potential given in \eqref{tachpotential1} as the 
following.
\bea
{\rm For} \quad R &\to& \infty, \qquad V(T(R)) = \tau_F\left(1 - \frac{Q_p}
{2R^{7-p}}\right) \simeq \tau_F\left(1 - \frac{Q_p}{2T^{7-p}}\right)
\label{Vinf}\\
{\rm and\,\,\,\, for} \quad R & \to& 0, \quad V(T(R)) = \tau_F 
\left(\frac{R^{\frac{7-p}{2}}}
{Q_p^{\frac{1}{2}}}\right) \simeq \tau_F Q_p^{\frac{1}{5-p}}
\left(-\frac{5-p}{2} T
\right)^{-\frac{7-p}{5-p}} \to 0 \label{Vzero}
\eea
We thus find that when F-string is far away ($R \gg Q_p^{\frac{1}{7-p}}$)
from the stack of D$p$-branes, the potential goes as $-1/T^{7-p} \simeq
-1/R^{7-p}$ which is the attractive gravitational potential in the 
$(9-p)$-dimensional transverse space of the D$p$-branes as expected. 
This is also consistent
with our earlier observation \eqref{ddotradg}. However, when the F-string
comes closer ($R \ll Q_p^{\frac{1}{7-p}}$) to the stack of D$p$-branes, 
the potential vanishes as power of the tachyon field.

In order to see how the pressure as given in \eqref{energypress} varies
with time, we need to solve the equation of motion of the radial mode $R$.
The equation of motion is given in \eqref{bounds} with the appropriate 
changes, namely,
\be
\dot R^2 = \frac{1}{H}\left(1 - \frac{1}{H}\left(\left(\frac{\tau_F V_F}
{E}\right)^2 + \frac{\ell^2}{R^2}\right)\right)\label{rdotnew}
\ee
with the harmonic function as given in \eqref{harmonicfn}. This equation
can be solved only when $R \ll Q_p^{\frac{1}{7-p}}$, i.e., for F-string
very close to the stack of D$p$-branes. Even for this case, we haven't been
able to solve the equation in a closed form for the non-zero angular
momentum. However, when the angular momentum is zero ($\ell =0$) 
eq.\eqref{rdotnew} can be solved and we get,
\be
t = \frac{2 Q_p^{\frac{1}{2}}}{5-p} R^{-\frac{5-p}{2}}\left(1 - F\left(
\frac{5-p}{2(7-p)},1;\frac{3}{2};\frac{Q_p}{(\frac{\tau_F V_F}{E})^2 +1} 
R^{-(7-p)}\right)\right)\label{time}
\ee 
where we have assumed the boundary condition that as $R \to 0$,
$t \to \infty$. Now as $R \to 0$, $F \to 0$ and so,
\be
t = \frac{2 Q_p^{\frac{1}{2}}}{5-p} R^{-\frac{5-p}{2}}
\ee
Inverting this relation we get,
\be
R = \left(\frac{5-p}{2 Q^{\frac{1}{2}}} t \right)^{-\frac{2}{5-p}}
\label{Rtime}
\ee
Substituting this into the expression of pressure given in \eqref{energypress}
we get,
\be
p = - \frac{\tau_F^2 V_F^2}{E} \left(\frac{5-p}{2}\right)^{
-\frac{2(7-p)}{5-p}} Q_p^{\frac{2}{5-p}} t^{-\frac{2(7-p)}{5-p}}
\ee
We, therefore, find that the pressure vanishes as a power of $t$ at late
times. This indicates that the system indeed evolves into the pressureless 
fluid or the tachyon matter state at late times as happens for the rolling 
tachyon solution of the non-BPS branes. We further note from 
\eqref{asymptotic1} and \eqref{Rtime}, that as $R \to 0$, 
\be
T(R \to 0) \to -t
\ee
So, the tachyon behaves as time, a typical of the rolling tachyon solution 
of the non-BPS branes \cite{Sen:2002qa}.

It should be noted here that the trajectory we have given in \eqref{Rtime}
is not valid at all times. This is because, for $p=1,2$ the dilaton behaves as
\be
e^{\phi} = g_s H^{\frac{3-p}{4}}
\ee
and for $R \ll Q_p^{\frac{1}{7-p}}$, the dilaton in general may not
remain small for all times. In order for the supergravity description
to remain valid we must ensure that $e^{\phi} \ll 1$. This implies a range
of $R$ for which the trajectory can be trusted and is given as,
\be
g_s^{\frac{4}{(7-p)(3-p)}} Q_p^{\frac{1}{7-p}} \ll R \ll Q_p^{\frac{1}{7-p}}
\ee
In terms of $t$, the corresponding range is 
\be
\frac{2}{5-p} g_s^{-\frac{2(5-p)}{(7-p)(3-p)}}Q_p^{\frac{1}{7-p}} \gg
t \gg \frac{2}{5-p} Q_p^{\frac{1}{7-p}}
\ee
This is the range of time for which the trajectory can be trusted. For small
string coupling $g_s$, this gives a sufficiently long time for which the 
trajectory remains reliable. Note also that in these cases the curvature
always remains small compared to the string scale.

For $p=3$, both the curvature and the string coupling can be made small. 
However, for $p=4$, even though the effective string coupling can be made
small, the curvature can become large and thereby invalidating the 
supergravity description. From the supergravity configuration \eqref{sugrasoln}
we compute the curvature for $p=4$ in the form ${\cal R} = 
(H^{1/2}R^2)^{-1/2}$ which must remain small compared to the string scale 
and so we have,
\be
\left(H^{-\frac{1}{2}}R^2\right)^{-\frac{1}{2}} \ll \frac{1}{\ell_s}
\ee
This implies
\be
R \gg \frac{\ell_s^4}{Q_4}
\ee
where $Q_4 = \pi N g_s \ell_s^3$. So, the supergravity description will remain
valid for
\be
\frac{\ell_s^4}{Q_4} \ll R \ll Q_4^{\frac{1}{3}}
\ee
In terms of $t$, the corresponding range is 
\be
\frac{2Q_4}{\ell_s^2} \gg t \gg 2 Q_4^{\frac{1}{3}}
\ee
For large $N$ this gives a sufficiently long time for which the trajectory
can be reliable.

Thus we observed that when a fundamental string moves in the vicinity of a
stack of coincident D$p$-branes it behaves like a non-BPS D-string in the
sense that it develops a tachyonic instability and eventually evolves into
a pressureless fluid state much like the tachyon matter of the rolling tachyon
solution of the non-BPS D-branes. The tachyon here has a geometric 
interpretation
in terms of the radial mode. At this point one would naturally be interested
to know what is the end point state of this geometric tachyon condensation.
From the space-time point of view, it is known that when F-string and the
stack of D$p$-branes are separated they break all the space-time supersymmetry
and so, it is not surprising that they develop a tachyonic
instability. However, when the tachyon rolls down from the initial homogeneous
configuration towards the minimum of the potential, F-string will evolve into
the pressureless fluid or tachyon matter. 
Usually in this situation the non-BPS D-branes evolving
into pressureless fluid is interpreted as the decay of the unstable D-branes
into the closed string radiation 
\cite{Larsen:2002wc,Okuda:2002yd,Lambert:2003zr,Gaiotto:2003rm}. 
Indeed, as argued for the case of motion of 
a D$p$-brane in the vicinity of a stack of coincident NS5-branes, F-string
also will lose almost all of its energy (decay) when it merges with the 
stack of D$p$-branes to form a bound state. This can be understood from the 
tension or the energy per unit
$p$-brane volume formula of an (F,D$p$) bound state which has the form 
\cite{Lu:1999uca,Schwarz:1995dk} 
\be
E_{N,M} =  T_0^p \sqrt{\frac{N^2}{g_s^2} + M^2}\label{energydensity}
\ee
where $T_0^p = [(2\pi)^p \ell_s^{p+1}]^{-1}$, $N$ is the number of
D$p$-branes and $M$ is the number of F-string per unit $(p-1)$-dimensional 
volume (along D$p$-brane), with $M$ and $N$ relatively prime.
It is clear from the energy formula that (F,D$p$) is a non-threshold
bound state and it is known to preserve half of the space-time supersymmetry. 
Suppose we consider a single F-string (per unit $(p-1)$-dimensional volume)
moving towards a stack of 
$N$ coincident D$p$-branes, where F-string lies parallel to the D$p$-branes.
From the above formula \eqref{energydensity}, we find that when the F-string 
gets bound to the stack of D$p$-branes, the system has an energy density
\be
E_{N,1} = T_0^p \sqrt{\frac{N^2}{g_s^2} + 1}
\ee
For the weak string coupling this can be expanded as
\be
E_{N,1} \simeq T_0^p \frac{N}{g_s} \left(1+\frac{g_s^2}{2N^2}\right) = 
T_0^p \left(\frac{N}{g_s} + \frac{g_s}{2N}\right)
\ee
Note that the second term which is the F-string contribution is almost
zero for weak string coupling and large $N$. Thus when the F-string approaches
the stack of D$p$-branes it loses almost all of its energy 
to form the bound state, much like the decay of unstable
D-branes to the pressureless tachyon matter. This is just a qualitative
argument, but to understand the detail dynamics of the bound state formation,
obviously, the supergravity description would not be sufficient and we need
the full string theoretic treatment.  
 
\section{Conclusion}

To summarize, in this paper after a brief review of proper acceleration
in both flat as well as curved space-time, 
we have given a systematic and detailed account of 
our observation that the origin of the geometric tachyon when a D$p$-brane
propagates in the close proximity of a stack of coincident NS5-branes is due
to the proper acceleration generated by the background dilaton field. We have seen this by
considering the motion of the D$p$-brane, described by the DBI action,
in terms of the motion of a 
relativistic point particle in the transverse space of the NS5-branes. The
acceleration is orthogonal to the proper velocity and is responsible for
changing the dynamics of the particle from that of a simple geodesic motion.
Using this we have seen that the motion of the particle in an appropriate
coordinate can be reduced to that of an inverted simple harmonic
oscillator. Even though the particle is not tachyonic in the conventional sense
(since $P^2 < 0$), effectively it behaves like a tachyon (as the motion
is hyperbolic) and the potential in which it moves has a maximum. In this
approach the maximum occurs at the location of the stack of
NS5-branes. However, since at that specific point the effective string coupling blows
up we can not trust the supergravity description. We have
computed the effective mass squared of the particle when it is close to
NS5-branes and found it to be negative. This, therefore, suggests the occurence
of tachyonic instability in the motion of a D$p$-brane in the vicinity of
NS5-branes.

We have extended our observations to the other dynamical systems, namely, the
motion of an F-string in the vicinity of a stack of coincident
D$p$-branes. The dynamics in this case is described by the Nambu-Goto
action. The Nambu-Goto action does not contain a dilaton prefactor
as the DBI action of a D$p$-brane, however, the induced metric of the
background D$p$-brane gives rise to a conformal factor which acts as a source
for the proper acceleration in this case. By using our previous observation we
showed that even in this case the proper acceleration is orthogonal to the
proper velocity and is responsible for the deviation of the dynamics of the
particle from its geodesic motion. Since $p=5$ case is quite different from
other $p$'s we have separately discussed these two cases. The motion of the
F-string in the background of a stack of D5-branes is S-dual to the motion of
a D-string in the background of a stack of NS5-branes and so, we follow
the discussion of $p=5$ case in close parallel with the D-string in NS5-brane
background. Here also, we showed that the acceleration is responsible for
modifying the motion of the radial mode of the particle in the vicinity of
D$p$-branes to that of an inverted
simple harmonic oscillator in an appropriate coordinate leading to the 
tachyonic instability. We have computed the mass squared of the particle
and found it to be negative. We have further studied the tachyonic instability
by mapping the Nambu-Goto action describing the dynamics of the F-string
to the tachyon effective action of the non-BPS D-string. We showed how the
tachyon potential behaves at various regimes by relating the radial mode to
the tachyon field. In particular, near the D5-branes we found that the
potential behaves like usual tachyon potential of a non-BPS brane. On the
other hand when the F-string is far away from the D$p$-branes the potential
behaves like a usual attractive $-1/R^2$ gravitational potential. But in this
approach we do not see the maximum of the potential where the instability 
begins to occur. We also
showed how the system evloves into pressureless fluid by solving the equation
of motion and it indeed falls off exponentially at late times like a non-BPS 
brane. 

Then we discussed the motion of an F-string in the vicinity of other
D$p$-branes. All our previous discussion goes through here, but the radial
equation of the particle motion is a bit involved in this case. Here we 
haven't been
able to show (in an appropriate coordinate) that the motion of the radial 
mode in the vicinity of the D$p$-branes can be modified to that of
an inverted simple harmonic oscillator. However, we showed that the potential
in which the particle moves has an extremum at $R=0$ and this extremum is a
point of inflexion, indicating an instability in the system. We have studied
the instability also by mapping the Nambu-Goto action to the tachyon effective
action of a non-BPS D-string. By relating the radial mode to the tachyon field
we have analyzed the behavior of the potential at various regimes. Here also
we found that when the F-string is far away from the stack of D$p$-branes
the potential behaves like $-1/R^{7-p}$ which is the attractive gravitational
potential as expected. However when the F-string is close to the stack of
D$p$-branes the potential vanishes as a power of the tachyon field unlike the
D5-brane background where the potential falls off to zero exponentially. We
have also solved the equation of motion of the radial mode when the F-string
moves in the vicinity of the stack of D$p$-branes with the zero angular 
momentum (since for non-zero angular momentum the equation can not be solved
in a closed form) and related the radial mode with the time. Using this we
have been able to show that the pressure of the system vanishes as power of
$t$ (unlike the D5-brane background where pressure falls off to zero
exponentially) at late times. This shows that the system indeed evolves into
the pressureless fluid (tachyon matter) like the rolling tachyon in non-BPS
branes. We have also shown that at late times the tachyon field behaves as
time a property typical of the rolling tachyon solution in non-BPS
branes. Further we have argued qualitatively that the (F,D$p$) bound state can
be understood as the result of this geometric tachyon condensation.       

\vspace{1cm}
     
{\bf ACKNOWLEDGMENTS}

\vspace{.5cm}

This work was supported in part by US DOE Grant number DE-FG 02-91ER40685.
Two of us (S.P. and S.R.) would like to thank the Department of Physics and
Astronomy of the University of Rochester, where most of this work was done,
for the warm hospitality.


\vspace{1cm}
        
\begin{thebibliography}{99}

\bibitem{Horowitz:1991cd}
  G.~T.~Horowitz and A.~Strominger,
  ``Black strings and P-branes,''
  Nucl.\ Phys.\  B {\bf 360}, 197 (1991).

\bibitem{Duff:1994an}
  M.~J.~Duff, R.~R.~Khuri and J.~X.~Lu,
  ``String solitons,''
  Phys.\ Rept.\  {\bf 259}, 213 (1995)
  [arXiv:hep-th/9412184].

\bibitem{Polchinski:1995mt}
  J.~Polchinski,
  ``Dirichlet-Branes and Ramond-Ramond Charges,''
  Phys.\ Rev.\ Lett.\  {\bf 75}, 4724 (1995)
  [arXiv:hep-th/9510017].

\bibitem{Sen:1999mg}
  A.~Sen,
  ``Non-BPS states and branes in string theory,''
  arXiv:hep-th/9904207.

\bibitem{Schwarz:1999vu}
  J.~H.~Schwarz,
  ``TASI lectures on non-BPS D-brane systems,''
  arXiv:hep-th/9908144.

\bibitem{Sen:1999md}
  A.~Sen,
  ``Supersymmetric world-volume action for non-BPS D-branes,''
  JHEP {\bf 9910}, 008 (1999)
  [arXiv:hep-th/9909062].

\bibitem{Garousi:2000tr}
  M.~R.~Garousi,
  ``Tachyon couplings on non-BPS D-branes and Dirac-Born-Infeld action,''
  Nucl.\ Phys.\  B {\bf 584}, 284 (2000)
  [arXiv:hep-th/0003122].

\bibitem{Bergshoeff:2000dq}
  E.~A.~Bergshoeff, M.~de Roo, T.~C.~de Wit, E.~Eyras and S.~Panda,
  ``T-duality and actions for non-BPS D-branes,''
  JHEP {\bf 0005}, 009 (2000)
  [arXiv:hep-th/0003221].

\bibitem{Kluson:2000iy}
  J.~Kluson,
  ``Proposal for non-BPS D-brane action,''
  Phys.\ Rev.\  D {\bf 62}, 126003 (2000)
  [arXiv:hep-th/0004106].

\bibitem{Sen:2004nf}
  A.~Sen,
  ``Tachyon dynamics in open string theory,''
  Int.\ J.\ Mod.\ Phys.\  A {\bf 20}, 5513 (2005)
  [arXiv:hep-th/0410103].

\bibitem{Leigh:1989jq}
  R.~G.~Leigh,
  ``Dirac-Born-Infeld Action From Dirichlet Sigma Model,''
  Mod.\ Phys.\ Lett.\ A {\bf 4}, 2767 (1989).

\bibitem{Schmidhuber:1996fy}
  C.~Schmidhuber,
  ``D-brane actions,''
  Nucl.\ Phys.\  B {\bf 467}, 146 (1996)
  [arXiv:hep-th/9601003].

\bibitem{Polchinski:1996na}
  J.~Polchinski,
  ``Lectures on D-branes,''
  arXiv:hep-th/9611050.


\bibitem{Kutasov:2004dj}
  D.~Kutasov,
  ``D-brane dynamics near NS5-branes,''
  arXiv:hep-th/0405058.

\bibitem{Sen:2002nu}
  A.~Sen,
  ``Rolling tachyon,''
  JHEP {\bf 0204}, 048 (2002)
  [arXiv:hep-th/0203211].

\bibitem{Sen:2002in}
  A.~Sen,
  ``Tachyon matter,''
  JHEP {\bf 0207}, 065 (2002)
  [arXiv:hep-th/0203265].

\bibitem{Sen:2002an}
  A.~Sen,
  ``Field theory of tachyon matter,''
  Mod.\ Phys.\ Lett.\  A {\bf 17}, 1797 (2002)
  [arXiv:hep-th/0204143].

\bibitem{Kutasov:2004ct}
  D.~Kutasov,
  ``A geometric interpretation of the open string tachyon,''
  arXiv:hep-th/0408073.

\bibitem{Sen:2007cz}
  A.~Sen,
  ``Geometric tachyon to universal open string tachyon,''
  JHEP {\bf 0705}, 035 (2007)
  [arXiv:hep-th/0703157].

\bibitem{Israel:2007zc}
  D.~Israel,
  ``Comments on geometric and universal open string tachyons near fivebranes,''
  JHEP {\bf 0704}, 085 (2007)
  [arXiv:hep-th/0703261].

\bibitem{Kluson:2007hb}
  J.~Kluson and K.~L.~Panigrahi,
  ``On the Universal Tachyon and Geometrical Tachyon,''
  JHEP {\bf 0706}, 015 (2007)
  [arXiv:0704.3013 [hep-th]].

\bibitem{Ghodsi:2004wn}
  A.~Ghodsi and A.~E.~Mosaffa,
  ``D-brane dynamics in RR deformation of NS5-branes background and tachyon
  cosmology,''
  Nucl.\ Phys.\  B {\bf 714}, 30 (2005)
  [arXiv:hep-th/0408015].

\bibitem{Papantonopoulos:2006eg}
  E.~Papantonopoulos, I.~Pappa and V.~Zamarias,
  ``Geometrical tachyon dynamics in the background of a bulk tachyon field,''
  JHEP {\bf 0605}, 038 (2006)
  [arXiv:hep-th/0601152].

\bibitem{Thomas:2005fu}
  S.~Thomas and J.~Ward,
  ``Inflation from geometrical tachyons,''
  Phys.\ Rev.\  D {\bf 72}, 083519 (2005)
  [arXiv:hep-th/0504226].

\bibitem{Panda:2005sg}
  S.~Panda, M.~Sami and S.~Tsujikawa,
  ``Inflation and dark energy arising from geometrical tachyons,''
  Phys.\ Rev.\  D {\bf 73}, 023515 (2006)
  [arXiv:hep-th/0510112].

\bibitem{Panigrahi:2007sq}
  K.~L.~Panigrahi and H.~Singh,
  ``Assisted Inflation from Geometric Tachyon,''
  JHEP {\bf 0711}, 017 (2007)
  [arXiv:0708.1679 [hep-th]].

\bibitem{Panigrahi:2008kg}
  K.~L.~Panigrahi and H.~Singh,
  ``Spinflation from Geometric Tachyon,''
  arXiv:0802.4230 [hep-th].

\bibitem{Das:2008af}
  A.~Das, S.~Panda and S.~Roy,
  ``Origin of the geometric tachyon,'' 
Phys.\ Rev.\ D {\bf 78}, 061901(R) (2008)
  [arXiv:0804.2863 [hep-th]].

\bibitem{Lu:1999uca}
  J.~X.~Lu and S.~Roy,
  ``Non-threshold (F,Dp) bound states,''
  Nucl.\ Phys.\  B {\bf 560}, 181 (1999)
  [arXiv:hep-th/9904129].

\bibitem{Schwarz:1995dk}
  J.~H.~Schwarz,
  ``An SL(2,Z) multiplet of type IIB superstrings,''
  Phys.\ Lett.\  B {\bf 360}, 13 (1995)
  [Erratum-ibid.\  B {\bf 364}, 252 (1995)]
  [arXiv:hep-th/9508143].

\bibitem{Elitzur:2000pq}
  S.~Elitzur, A.~Giveon, D.~Kutasov, E.~Rabinovici and G.~Sarkissian,
  ``D-branes in the background of NS fivebranes,''
  JHEP {\bf 0008}, 046 (2000)
  [arXiv:hep-th/0005052].

\bibitem{Pelc:2000kb}
  O.~Pelc,
  ``On the quantization constraints for a D3 brane in the geometry of NS5
  branes,''
  JHEP {\bf 0008}, 030 (2000)
  [arXiv:hep-th/0007100].

\bibitem{Ribault:2003sg}
  S.~Ribault,
  ``D3-branes in NS5-branes backgrounds,''
  JHEP {\bf 0302}, 044 (2003)
  [arXiv:hep-th/0301092].

\bibitem{Lu:1998vh}
  J.~X.~Lu and S.~Roy,
  ``An SL(2,Z) multiplet of type IIB super five-branes,''
  Phys.\ Lett.\  B {\bf 428}, 289 (1998)
  [arXiv:hep-th/9802080].

\bibitem{Mitra:2000wr}
  I.~Mitra and S.~Roy,
  ``(NS5, Dp) and (NS5, D(p+2), Dp) bound states of type IIB and type IIA
  string theories,''
  JHEP {\bf 0102}, 026 (2001)
  [arXiv:hep-th/0011236].

\bibitem{Alishahiha:2000qv}
  M.~Alishahiha and Y.~Oz,
  ``Supergravity and 'new' six-dimensional gauge theories,''
  Phys.\ Lett.\  B {\bf 495}, 418 (2000)
  [arXiv:hep-th/0008172].

\bibitem{Duff:1990wv}
  M.~J.~Duff and J.~X.~Lu,
  ``Elementary five-brane solutions of D = 10 supergravity,''
  Nucl.\ Phys.\  B {\bf 354}, 141 (1991).

\bibitem{Callan:1991dj}
  C.~G.~.~Callan, J.~A.~Harvey and A.~Strominger,
  ``World sheet approach to heterotic instantons and solitons,''
  Nucl.\ Phys.\  B {\bf 359}, 611 (1991).

\bibitem{Rindler} W. Rindler, Phys. Rev. {\bf 119}, 2082 (1960); 
Amer. Jour. of Phys. {\bf 34}, 1174 (1966).

\bibitem{Panigrahi:2004qr}
  K.~L.~Panigrahi,
  ``D-brane dynamics in Dp-brane background,''
  Phys.\ Lett.\  B {\bf 601}, 64 (2004)
  [arXiv:hep-th/0407134].

\bibitem{Sen:2002qa}
  A.~Sen,
  ``Time and tachyon,''
  Int.\ J.\ Mod.\ Phys.\  A {\bf 18}, 4869 (2003)
  [arXiv:hep-th/0209122].

\bibitem{Larsen:2002wc}
  F.~Larsen, A.~Naqvi and S.~Terashima,
  ``Rolling tachyons and decaying branes,''
  JHEP {\bf 0302}, 039 (2003)
  [arXiv:hep-th/0212248].

\bibitem{Okuda:2002yd}
  T.~Okuda and S.~Sugimoto,
  ``Coupling of rolling tachyon to closed strings,''
  Nucl.\ Phys.\  B {\bf 647}, 101 (2002)
  [arXiv:hep-th/0208196].

\bibitem{Lambert:2003zr}
  N.~D.~Lambert, H.~Liu and J.~M.~Maldacena,
  ``Closed strings from decaying D-branes,''
  JHEP {\bf 0703}, 014 (2007)
  [arXiv:hep-th/0303139].

\bibitem{Gaiotto:2003rm}
  D.~Gaiotto, N.~Itzhaki and L.~Rastelli,
  ``Closed strings as imaginary D-branes,''
  Nucl.\ Phys.\  B {\bf 688}, 70 (2004)
  [arXiv:hep-th/0304192].
























\end{thebibliography}
\end{document}